\font\big=cmr17
\font\title=cmbx12
\vglue 1 cm
\centerline{\big{CSL Collapse Model And Spontaneous Radiation: An Update}}
\vskip 1cm
\centerline {Philip Pearle and James Ring} 
\centerline{\it{Department of Physics, Hamilton College, Clinton, N.Y.,13323}}
\bigskip
\centerline {Juan I. Collar}
\centerline{\it{EP Division, CERN CH-1211}}
\bigskip
\centerline{Frank T. Avignone III} 
\centerline{\it{Department of Physics and Astronomy, 
University of South Carolina, Columbia South Carolina 29208}}

\vskip 1cm
\centerline{\tenbf{Abstract}}

{\tenrm A brief review is given of the Continuous Spontaneous Localization (CSL) 
model in which a classical field interacts with quantized 
particles to cause dynamical wavefunction collapse.  
One of the model's predictions is that particles 
``spontaneously" gain energy at a slow rate.  
When applied to the excitation of a nucleon in a Ge nucleus, 
it is shown how   
a limit on the relative collapse rates of neutron and proton could be obtained, 
and a rough estimate is made from data.        
When applied to the spontaneous excitation of 1s electrons in Ge, by a more     
detailed analysis of more accurate data than 
previously given, an updated limit is obtained on 
the relative collapse rates of the electron and proton, suggesting that 
the coupling of the field to electrons and nucleons is mass proportional.}

\vskip 1cm
\leftline {\title 1. Introduction}
\bigskip

	It is appropriate to discuss 
comparison of experiment to a theory with fundamental pretensions in a volume 
dedicated to Dan Greenberger whose own theoretical work of a fundamental nature 
has seldom been far from testability.  

	In standard quantum theory (SQT) the statevector evolves in two ways. One 
evolution procedes smoothly via Schrodinger's equation. The other is the 
abrupt (and ill defined)  ``collapse" of the statevector. This is the replacement 
of a statevector equal to 
a sum of vectors (each describing a different outcome of an ``experiment") by one 
vector in the sum. One might guess that this dual evolution is 
indicative of a fundamental deficiency in present day physics.   
In the hope of finding new physics, one may begin by trying to modify Schrodinger's 
equation so that the statevector undergoes only a 
smooth evolution, giving both the usual quantum behavior and the collapse behavior.

	This program, begun three decades ago,$^{1,2}$ received a crucial impetus 
one decade ago from the work of 
Ghirardi, Rimini and Weber (GRW)$^{3}$ and has evolved into 
the Continuous Spontaneous Localization (CSL) model.$^{4}$  At present, this is the 
only fully developed nonrelativistic collapse model, with definite predictions 
applicable to any nonrelativistic experimental situation.  

	In the CSL model, a term which depends upon a 
randomly fluctuating field $w({\bf x},t)$ is added to Schrodinger's 
equation.  (The physical nature of this 
field is unspecified, but metric fluctuations,$^{5}$ possibly with a 
tachyonic spectrum,$^{6}$ have been suggested.)  The probabilistic behavior of Nature 
is explained as due to our lack of control of the field $w$.
When an experiment is under way,the particles 
in the system+apparatus interact with the particular sample field $w$ that is present, causing  
a rapid evolution of the statevector to one of the alternative outcomes of the experiment.  
A different sample field leads to a different outcome.  CSL also specifies the probability that 
a particular sample field $w({\bf x},t)$ actually occurs.  When all 
possible fields are taken into account, together with their probabilities, 
the result is that each outcome occurs with (essentially --- see next paragraph) the 
probability predicted by SQT. Thus two relations, the  
Modified Schrodinger Equation and the Probability Rule constitute the CSL model.  

	As with any modification of SQT, one expects---and hopes---for certain 
specially designed experiments where SQT and CSL lead to different predictions, 
making tests possible. For example, one such test,  
presently not practicable, is a two slit interference experiment with a
sufficiently large bound state object.  Once the two wavepackets for the object 
leave the slits, SQT says that their amplitudes will never change 
so that interference is possible at any time.  CSL says that the amplitudes will 
fluctuate and, after a long enough wait, eventually one of the packets will 
become negligible in amplitude, giving no interference pattern.  
Such an interference experiment with e.g., 
90\AA\  diameter drops of mercury$^{7}$ over a time interval of seconds 
could provide such a test.  However, this is a difficult experiment.  For 
example, it is hard to prevent the two packets from  
being put into different angular momentum eigenstates by interaction with the 
environment, and then they would not interfere for this reason.$^{8}$ 

	The tests that are most practicable at present stem from 
the consequence of CSL that the collapse process imparts energy to particles.   
(One may think of this energy as provided by the field $w$.)  The reason is that 
collapse entails the narrowing of wavefunctions.  By the uncertainty principle, 
this leads to an increased momentum spread, and thus to an increased energy.  
Thus, any bound ground state of, e.g., atoms or nucleii, will be excited 
by the collapse part of the Schrodinger equation.$^{3,4,9}$  The usual part of the 
Schrodinger equation will describe the radiation emitted as the system returns to the 
ground state. Also, a free charged particle is ``shaken" by the 
field $w$, and so it will radiate$^{10}$.  Similarly,  
the quarks inside a proton should be excited, and the proton should radiate mesons.$^{11}$  
Thus, a signature of CSL is that matter should emit ``spontaneous 
radiation." In section 3, various radiation rates are given.

	It is interesting that, at this time, 
quite a number of low noise experiments are being undertaken to look for radiation 
appearing in an apparatus for a variety of reasons, e.g., because of collisions
with purported Dark Matter. Some of these experiments are sensitive 
enough to provide useful constraints on the parameters of CSL. 
 
	Some experiments which look at radiation 
appearing in a slab of Germanium are described in section 4.   
In section 5 we show how data from one such experiment, ``Rico Grande," applied to 
the spontaneous excitation of a proton in a Ge nucleus, can 
provide a limit on the relative collapse rates of neutron and proton.   
Only a rough estimate is given because greater precision requires a more careful calculation of 
Ge nuclear dipole matrix elements than we are prepared to give here.
   
	In section 6 we consider the spontaneous ionization rate of a 1s electron in 
a Ge atom. In a previous paper,$^{12}$ it was argued that the upper limit on this rate 
given by a single data point from the ``TWIN" experiment$^{13}$ suggested 
that the coupling of the field $w$ to an electron or nucleon 
(here assumed to be the same for neutron and proton) is 
proportional to the particle's mass, supporting 
previous proposals that there is a connection between 
gravity and collapse$^{5,14,15}$.  Unfortunately it subsequently turned out that 
the data point was inaccurate (see section 4).  However, the more complete 
analysis on ``COSME" data presented here gives essentially the previous result.    
 
	Eventually, from such experiments, one may hope 
that the collapse rate parameter of CSL will either be constrained to be so
small that the model will be ruled out---or that spontaneous radiation from 
collapse will actually be observed! 

\vskip 1cm
\leftline {\title 2. CSL}
\bigskip

	Underlying CSL are two mechanisms. One is the  
Gambler's Ruin mechanism. This explains how the random process 
embodied in the noise $w$ produces the probabilities of SQT 
for the collapsed states.$^{16}$ The other is 
the GRW ``hitting" mechanism. It allows collapse to occur rapidly, 
for macroscopic objects, to states which we  
see around us (localized objects), while
microscopic objects are scarcely affected.$^{3}$  

	Here is the Gambler's Ruin analogy.  
Suppose, at the beginning of a game, gambler 1 (2) starts with $d_{1}(0)$ ($d_{2}(0)$) 
dollars, and $d_{1}(0)+d_{2}(0)=100$.  This is to be analogous to 
the initial statevector $|\psi (0)>=a_{1}(0)|1>+a_{2}(0)|2>$, with the 
correspondence $d_{i}(0)/100\rightarrow |a_{i}(0)|^{2}$. The gamblers toss a coin 
(analogous to the fluctuating field $w$) and, depending on the result, 
one gives a dollar to the other.  As the game proceeds, the $d_{i}(t)$ fluctuate,       
just as do the squared amplitudes $|a_{i}(t)|^{2}$.    
One gambler finally wins all the money and the game stops, with e.g., 
gambler 2 winning with probability $d_{2}(0)/100$. 
Precisely analogously, collapse finally occurs, 
e.g., with $|\psi (t)>\rightarrow 0|1>+1|2>$ with probability $|a_{2}(0)|^{2}$.  
This is, of course, the 
probability predicted by SQT of the outcome $|2>$ if $|1>$ and $|2>$ represent two states of 
an apparatus.  

	The GRW model postulates a physical process which produces a sudden random change (``hit") 
of a many-particle wavefunction:   
the wavefunction is multiplied by a gaussian function 
$\exp-({\bf x}_{n}-{\bf z})^{2}/2a^{2}$, where ${\bf x}_{n}$
is the position coordinate of the nth particle. The center of the gaussian, ${\bf z}$, is 
chosen according to a Probability Rule which depends in a certain way upon the 
wavefunction, making it most likely that ${\bf z}$ is located where 
the wavefunction is largest.  The effect of a hit is to narrow to 
width $a$ the part of the wavefunction which depends upon the nth particle: 
GRW chose the mesoscopic length $a\approx 10^{-5}$ cm.  A hit on one particle occurs 
at a slow rate $\lambda$: GRW chose $\lambda\approx 10^{-16}$sec$^{-1}\approx$ 
once in 300 million years.  But, each particle is 
equally likely to be hit, so a hit on an N particle 
object occurs rapidly, on average in $1/\lambda N$sec.  The wavefunction of an 
$N$ particle object in a superposition of states, each describing 
the object in a different place, has the particles entangled in such a way that 
one hit on one such particle causes the wavefunction 
to collapse in $1/\lambda N$sec to one of the states in the superposition.   
This explains how macroscopic objects are always observed as localized.  
(A defect of this model is that the 
(anti) symmetry of the wavefunction is destroyed by the hitting process.) 
	
		CSL may be thought of as embodying a continuous hitting process. 
A hit occurs every $\Delta t$ sec, 
but the hit wavefunction is 
multiplied by $\Delta t$ and added to 
the original wavefunction.  Thus the wavefunction shape 
fluctuates gradually (not suddenly as in GRW's model), 
and the gambler's ruin dynamics is obtained. The 
(anti) symmetry of the wavefunction is properly preserved in CSL.  

	The CSL modified Schrodinger equation is 
	
$${d|\psi,t>_w\over dt}= -iH|\psi,t>_w - 
{1\over 4\lambda}\int d{\bf x}[w({\bf x},t)-2\lambda A({\bf x})]^2|\psi,t>_w\eqno (2.1)$$

\noindent Given an arbitrary field $w({\bf x},t)$, one may solve (2.1) to 
find how the statevector evolves under its influence.  
Operator A({\bf x}) is

	$$A({\bf x})\equiv \sum _{\alpha} g_{\alpha}
	{1\over (\pi a^{2})^{3\over 4}}\int d{\bf z}N_{\alpha}({\bf z}) 
	e^{\displaystyle -{({\bf x}-{\bf z})^{2}
	\over 2a^{2}}}\eqno (2.1a)$$

\noindent  The integral in (2.1a) is essentially the 
number of particles in a sphere of diameter $a$ centered around ${\bf x}$. 
$N_{\alpha}({\bf z})$ is the number density operator for particles of type $\alpha$, e.g., 
electrons , protons and neutrons, and 
$\lambda g_{\alpha}^{2}$ is their one-particle collapse rate (see Eq. (2.5) below).
 
	Thus, the parameters which characterize 
CSL are $a$, $\lambda$ and the ratios of the $g_{\alpha}$'s, e.g., for ordinary matter, $g_{n}/g_{p}$ 
and $g_{e}/g_{p}$ where the subscripts $p$, $n$, and $e$ refer to the proton, neutron and electron. 
With no loss of generality we may take $g_{p}=1$, so $\lambda$ is 
an individual proton's collapse rate. 
 
	CSL requires a second equation, giving the probability density $P(w)$ functional that the field 
$w({\bf x},t)$ occurs:

	$$P(w)=\thinspace_{w}\negthinspace\negthinspace<\psi,t|\psi,t>_w\eqno(2.2)$$
	
\noindent Because the evolution (2.1) is nonunitary, the norm of the statevector $|\psi,t>_w$ 
changes with time. Eq.(2.2) says that the most probable fields to occur 
are those which lead to statevectors of largest norm. 

	Eqs.(2.1) and (2.2) ensure that an initial statevector, which is in a superposition of 
states of different particle number density, evolves toward one of these 
states for each probable $w$---and that the ensemble of all evolutions is such that 
each final state occurs with (essentially) the SQT probability.  
However, we shall not discuss here how Eqs.(2.1), (2.2) lead to collapse for an individual 
statevector.  Instead we shall go right to the appropriate object 
for discussing experimental predictions, the density matrix $\rho$.  
The evolution equation for 
the density matrix whch follows from Eqs.(2.1), (2.2) can be shown to be $^{4}$

$$\eqalign{&{\partial <x|\rho(t)|x'>\over\partial t}= -i<x|[H,\rho(t)]|x'>\cr
&\quad -{\lambda\over2}\sum^{N}_{j=1}\sum^{N}_{k=1}
g_{\alpha(j)}g_{\alpha(k)}[\Phi({\bf x}_j-{\bf x}_k)+\Phi({\bf x}'_j-{\bf x}'_k)
-2\Phi({\bf x}_j-{\bf x}'_k)]<x|\rho(t)|x'>\cr}\eqno(2.3)$$

\noindent where $|x>=| {\bf x}_1,{\bf x}_2,\dots>$ is the position eigenstate
for all particles  and

$$\Phi({\bf z})\equiv e^{\displaystyle-{{\bf z}^{2}\over4a^{2}}}\eqno(2.4)$$ 

	As a simple example of how Eq. (2.3) works, set $H=0$ so as to 
concentrate on the collapse dynamics alone, and consider a clump of particles 
of type $\alpha$ in a superposed state.  That is, let the 
initial state be $a_{1}(0)|1>+a_{2}(0)|1>$, where $|1>$ and $|2>$ each describe 
N particles of type $\alpha$ in a localized state with dimensions $<<a$, but 
with centers of mass of the two states at a distance $>>a$ apart.  
Then $\Phi({\bf x}_j-{\bf x}_k)\approx 1$ if $ {\bf x}_j$, ${\bf x}_k$ are both located 
in region 1 (or both in 2) and $\Phi({\bf x}_j-{\bf x}'_k)\approx 0$ if 
${\bf x}_j$,  ${\bf x}'_k$ are located in regions 1 and 2 respectively. Therefore, 
Eq.(2.3) yields 

$${\partial <1|\rho(t)|2>\over\partial t}=-{\lambda g_{\alpha}^{2}\over 2} 	
	\sum^{N}_{j=1}\sum^{N}_{k=1}(1+1-2\cdot 0)<1|\rho(t)|2>
	=-\lambda g_{\alpha}^{2}N^{2}<1|\rho(t)|2>\eqno(2.5)$$
	
\noindent showing that the off-diagonal elements of $\rho$ decay 
at the rate $\lambda N^{2}$.  
This illustrates how the collapse rate is large for a superposition of states describing a 
large number of particles in different locations.

\vskip 1cm
\leftline {\title 3.  Excitation Rate Predictions }
\bigskip 
 
	As mentioned in section 1, a byproduct of the collapse process is 
that particles gain energy.  It is easy to show, using Eq.(2.3), that 
the average total energy ${\bar H}(t)\equiv$Tr$H\rho (t)$ 
($H=\sum_{i=1}^{N} {\bf p}_{i}^{2}/2M_{i}+V({\bf x}_{1}\dots {\bf x}_{N})$) increases 
according to 

$${d{\bar H}(t)\over dt}={3\lambda\over 2}\sum_{j=1}^{N}g_{\alpha(j)}^{2}{\hbar ^{2}
\over 2M_{j}a^{2}}\eqno(3.1)$$

\noindent Assuming $g_{\alpha}=1$ for all particles, and using the GRW values for 
$\lambda$ and $a$, then $10^{24}$ nucleons gain $\approx .3$ eV/sec and 
$10^{24}$ electrons gain $\approx 600$ eV/sec.  This is quite small, corresponding to a 
temperature increase over the age of the universe of $\approx .001^{\circ}$K and $2^{\circ}$K 
respectively. (The low particle density in the universe assures that 
the effect of this increased energy on the the cosmic radiation bath 
is negligible). If we take $10^{24}$ electrons as roughly the number in a cc. of condensed matter, 
this corresponds to an energy increase of about $10^{-15}$ joules/sec, 
which is close to the experimentally detectable lower bound 
by present day bolometric measurements.$^{17}$  This is much 
less sensitive than the experiment discussed here.   

 However, while (3.1) gives the {\it average} behavior, 
there are infrequent but large energy fluctuations. 

	The energy increase in Eq.(3.1) is the sum of increased internal energy and of 
increased center of mass energy 
$H_{cm}\equiv {\bf P}_{cm}^{2}/2M +V_{cm}({\bf Q}$) (${\bf P}_{cm}\equiv \sum_{j=1}^{N}
{\bf p}_{j}$, $M\equiv \sum_{j=1}^{N} M_{j}$, 
${\bf Q}\equiv \sum_{j=1}^{N}M_{j}{\bf x}_{j}/M$).  From Eq.(2.3) we find
 
$$\eqalign{{d{\bar H}_{cm}(t)\over dt}&={3\lambda\hbar ^{2}\over 4a^{2}M}
\sum_{j=1}^{N}\sum_{k=1}^{N}g_{\alpha(j)}g_{\alpha(k)}\hbox{Tr}
\biggl\{\biggl[1-{({\bf x}_{j}-{\bf x}_{k})^{2}\over 6a^{2}}\biggr]
e^{-{({\bf x}_{j}-{\bf x}_{k})^{2}\over 4a^{2}}}\rho (t)\biggr\}\cr
&\qquad\qquad\qquad={3\lambda\hbar ^{2}\over 4a^{2}M}
\biggl[\sum_{j=1}^{N}g_{\alpha(j)}\biggr]^{2}-o(a^{-4})\cr}\eqno(3.1a,b)$$
 
\noindent where the first term in the expansion (3.1b) dominates if 
the system under consideration, like an atom or nucleus, has dimensions $<<a$.  The 
condition for the total energy increase to be completely due to the 
center of mass energy increase 
to order $a^{-2}$, i.e., for there to be no internal excitation to this order, 
is found by equating (3.1) to (3.1b):

$$0=\sum_{j=1}^{N}g_{\alpha(j)}^{2}/M_{j}-\biggl[\sum_{j=1}^{N}g_{\alpha(j)}\biggr]^{2}/M=
(2M)^{-1}\sum_{j=1}^{N}\sum_{k=1}^{N}
\biggl[g_{\alpha(j)}\sqrt{M_{k}/M_{j}}-g_{\alpha(k)}\sqrt{M_{j}/M_{k}}\biggr]^{2}$$

\noindent i.e., if $g_{\alpha(j)}=CM_{j}$.  Therefore, for such mass--proportionality of the 
coupling constants, the internal energy does not increase to order $ a^{-2}$. 
Moreover, the leading term in the internal energy increase, 
proportional to $ a^{-4}$, is
compensated by an identical decrease in the center of mass energy since, by (3.1), 
the total energy increase vanishes to order $a^{-4}$ and higher.   
	
		If the coupling constants are not mass--proportional, 
the internal excitation rate $\sim a^{-2}$ is found as follows. Consider a 
transition from a state $|\psi>|\chi>$ to a state 
$|\phi>|\chi '>$, where $|\psi>$ is an initial bound state, $|\phi>$ is an orthogonal 
final state, $|\chi>$ is an initial state of the center of mass and $|\chi '>$ is 
an arbitrary final state of the center of mass. The probability per second of a transition from 
$|\psi>$ to $|\phi>$, regardless of the final center of mass state is 
$\dot P \equiv \sum_{|\chi '>}<\chi '|<\phi|\dot\rho (0)|\phi>|\chi '>$, where 
$\rho (0)=|\chi>|\psi><\psi|<\chi|$.  Expansion of  
Eq. (2.3) to first order in (dimension of system/$a$)$^{2}$ yields$^{11}$  

	$${\dot P}_{1}={\lambda\over 2a^{2}}
	<\phi|{\bf R}|\psi>{\bf \cdot}<\psi|{\bf R}|\phi>\eqno (3.2)$$  

\noindent where ${\bf R}\equiv \sum_{j=1}^{N}g_{\alpha(j)}{\bf R}_{j}$,  
${\bf R}_{j}\equiv {\bf x}_{j}-{\bf Q}$. 

	It is the predictions of Eq.(3.2) that we shall test in this paper. It is worth remarking that 
the matrix element in Eq.(3.2) involving a charged particle's ${\bf R}_{j}$  
is the same as that involved in describing an electric dipole 
transition between $|\psi>$ and $|\phi>$. Thus one can evaluate this contribution to 
(3.2) either by calculating 
the relevant matrix element or by expressing it in 
terms of measurable transition rates. Indeed, as we shall see in sections 5 and 6, 
using ${\bf R}\equiv 0$ if $g_{\alpha}\sim M_{\alpha}$, it is possible to express 
the matrix elements of one type of particles in terms of another type 
so one need only calculate or measure the matrix 
elements of the excited particle type to evaluate (3.2).  
  
Incidentally, by summing Eq.(3.2) over all states $|\phi>$ orthogonal to $|\psi>$, we obtain 
${\dot P}_{1}^{T}$, the total probability/sec for excitation of $|\psi>$: 

$${\dot P}_{1}^{T}={\lambda\over 2a^{2}}<\psi|[{\bf R}-<\psi|{\bf R}|\psi>]^{2}|\psi>\eqno (3.3)$$
	  
	Assume the GRW values for 
$\lambda$ and $a$.  For an atomic electron undergoing spontaneous excitation from, e.g., 
the 1s state of an atom with atomic number 
$Z$ to a higher energy state, bound or free, the order of magnitude of 
${\dot P}_{1}$ is 
$\approx g_{e}^{2}10^{-23}/Z^{2}$sec$^{-1}$. 
For, e.g., a proton in an outer shell of a nucleus of 
mass number $A$ it is $\approx g_{p}^{2}10^{-32}A^{2/3}$sec$^{-1}$. 
With such rates, the 1s electrons in $10^{24}$ such atoms would 
be expected to provide $\approx g_{e}^{2}10/Z^{2}$ 
photon pulses each second while each spontaneously excited 
proton in $10^{24}$ such nucleii would 
be expected to provide $\approx g_{p}^{2}.3A^{2/3}$ gammas each year. This large difference in rates 
explains why we are able to 
obtain good experimental limits on the electron's coupling constant $g_{e}$ in section 6, but 
not on the neutron's coupling constant $g_{n}$ in section 5.  

	Although we shall only apply Eq. (3.2) in our data analysis, for 
completeness we include Eqs. (3.4), (3.5) below which could be appled if more accurate data
becomes available.  We note again that if $g_{\alpha}$ is 
mass--proportional (i.e., if for protons $g_{p}=1$, then for neutrons 
$g_{n}\approx 1.001$ and for electrons $g_{e}\approx .00054$), 
then ${\bf R}\equiv 0$, and therefore (3.2) vanishes. We should 
then need ${\dot P}$ to order $a^{-4}$:

$${\dot P}_{2}= {\lambda\over 16a^{4}}[ |<\phi|S|\psi>|^{2} + 
2\sum_{m=1}^{3}\sum_{n=1}^{3} |<\phi|S^{mn}|\psi>|^{2}]\eqno(3.4)$$

\noindent where $S^{mn}\equiv \sum_{j=1}^{N}g_{\alpha(j)}({\bf R}_{j})^{m}({\bf R}_{j})^{n}$,
$S\equiv \sum_{n=1}^{3}S^{nn}$. If  ${\bf R}_{j}$ corresponds to a charged particle, 
its matrix elements here 
describe electric monopole or quadrupole transitions. These matrix elements are smaller 
than the corresponding matrix elements in Eq.(3.2) by the factor 
(size of bound state/$a$)$^{2}$.   

	Fu$^{10}$ has considered the spontaneous electromagnetic radiation of a free charged 
particle in CSL, obtaining the probability/sec/energy of radiating a photon of 
energy $E=\hbar k$:

$$ {d{\dot P}(E)\over dE}=g_{\alpha}^{2}{\lambda\over 4\pi^{2}}{e^{2}\over\hbar c}
\Bigl( {\hbar/M_{\alpha}c\over a}\Bigr)^{2}{1\over E}\approx 
{3\cdot 10^{-31}\over E\hbox{\thinspace\thinspace in keV}}
\hbox {\thinspace\thinspace counts/sec/keV}\eqno (3.5a,b)$$ 

\noindent (the infrared divergence is treated as usual).  Eq.(3.5b) gives the 
rate for an electron with $g_{e}=1$.  This radiation 
rate (3.5) for free particles is smaller than the excitation rate 
(3.2) for bound particles by the factor $e^{2}/\hbar c\approx 1/137$ and 
by the replacement of (bound state size/$a$)$^{2}$ by (Compton wavelength/$a$)$^{2}$. 
We note, with mass-proportionality, that the spontaneous radiation 
rate for free electrons is the same as for free protons, on account of 
the factor $(g_{\alpha}/M_{\alpha})^{2}$ in (3.5a).

	This completes our collection of equations giving CSL spontaneous 
excitation rates. 

	We now consider the spontaneous excitation of valence nucleons in a 
Germanium nucleus and the spontaneous ionization of 1s electrons in a Germanium atom, 
using data from 
what are, at present, the lowest noise relevant experiments.     

\vskip 1cm
\leftline {\title 4. Experiments.}
\bigskip  

	The data used in the analysis$^{18}$ of the atomic 
excitation comes from a small (253 g) p-type coaxial HPGe
crystal, ``COSME", built specifically for a Dark Matter search$^{19}$.   It
features, at only 1.6 keV, the (so-far) lowest energy threshold of 
any detector dedicated to
such searches, and has as well an excellent resolution of 0.43 keV
(FWHM) at 10.3 keV. Special low-radioactivity measures were taken, such as
mounting the detector on a specially-designed electroformed copper
cryostat, shielding of electronic components close to the crystal with
450-yr-old lead, and use of a 2000-yr-old roman lead layer in the innermost
part of the shielding. Additional photon, neutron and vibrational shielding
were used. 

	The detector set-up was installed in the Canfranc-1
underground laboratory in the Spanish Pyrenees, at a depth of 675 meters of
water equivalent. The microphonic component characteristic of very
low-threshold detectors, extending up to $\approx 15$keV, was filtered-out using
specially-developed techniques $^{20}$. While the low-energy background level
was slightly higher than that in the ``TWIN' detectors $^{13}$, the improved
resolution ---typically inversely proportional to the mass of the crystal---
allows one to impose more stringent limits on a sharply defined signal that
might be buried in an otherwise featureless background, as is the case for
the emitted radiation in CSL. 

	At this point a remark is in order:
unfortunately the TWIN data used to extract CSL limits on $g_{e}$ in reference 12 belonged to
a preliminary set coming from un-amplified digitized pulses. Later
comparison with the spectrum collected with a multichannel analyzer showed
that this earlier data was corrupted at energies below $\approx 200$ keV, i.e., a
large fraction of events were not recorded. This faulty set was not
used in other TWIN results, namely for double-beta decay $^{13}$, Dark Matter
$^{21}$ or electron half-life $^{22}$.  In reference 12, only one data point was used, the 
(erroneous) rate .049 counts/keV/kg/day at 11 keV (where the highest predicted value of 
spontaneous radiation occurs) to set an upper limit on $g_{e}$.  In 
the present work, it is necessary to perform a Chi$^{2}$ analysis of a fit to the whole 
predicted shape of the spontaneous radiation, in order to obtain a 
limit comparable to the (erroneous) limit of reference 12. 

	The data used in the analysis of the nuclear excitation comes from the ``Rico Grande" 
crystals, which are part of the IGEX ensemble of large enriched
germanium detectors$^{23}$, dedicated to searching for neutrinoless
double-beta decay. The two detectors from which these data were extracted
have a fiducial mass of $\approx$2 kg each and are enriched to 86\% Ge$_{76}$ and 14\%
Ge$_{74}$. As a result, the prevailing source of background 
in the energy neighborhood of interest here is this two-neutrino double-beta
decay from Ge$_{76}$ which, however, cuts off at an energy below 
the region employed in the analysis in section 5. 
At the time of collection of the present data, the Ricos
were operated in the Homestake mine in similar conditions to TWIN.

\vskip 1cm
\leftline {\title 5. Constraint on $g_{n}$/$g_{p}$?}
\bigskip

	We now apply Eq. (3.2) 
to the spontaneous excitation rate of a single proton or neutron in a Ge nucleus. 

The Jparity for the ground state of Ge is 0+ for the even-even nuclides 
Ge$_{70}$ (20.6\%), Ge$_{72}$ (27.4\%), Ge$_{74}$ (36.7\%) and Ge$_{76}$ (7.7\%)),  
and it is 9/2+ for Ge$_{73}$ (7.7\%). 
Spontaneous excitation is predicted (the matrix elements are nonvanishing) 
for a transition from the ground state to 1- states  
in the case of the even-even nuclides and to 
11/2-, 9/2- or 7/2- states in Ge$_{73}$. From the point of 
view of the shell model, among other possibilities, a proton or neutron 
can be excited from its ground state valence level 
to a higher energy state.   

	The return of a proton from the excited state to ground could be direct,  
via an electric dipole transition. It could also proceed indirectly, 
through intermediate states or internal conversion and, in the case of 
a neutron it must proceed indirectly, through magnetic transitions.   
In any case, the lifetimes are in most cases so short that 
a photon pulse would rapidly appear at the energy difference of the two states. 
Therefore, by looking at the data for a signature peak 
of instrumental resolution width at the expected energy, one may hope to 
observe the radiation resulting from 
these spontaneous transitions or at least get an upper limit on their rate.

	Not only does the matrix element of ${\bf R}_{i}$ for the excited particle 
not vanish, but the matrix element of ${\bf R}_{i}$ for the other particles also does not vanish 
due to their dependence on the center of mass operator.  We can relate the 
matrix element of the protons to that of the neutrons by using 
$\Sigma_{j}M_{j}{\bf R}_{j}\equiv 0$, which implies that 
 
$$\Sigma_{i=1}^{A-Z}{\bf R}_{ni}=-(M_{p}/M_{n})\Sigma_{i=1}^Z{\bf R}_{pi}\eqno (5.1a)$$

\noindent (we neglect the electron contribution of o$(M_{e}/(M_{n})$) so, setting
$M_{p}/M_{n}=1$, the matrix element in (3.2) is

$$<\phi|g_{p}\Sigma_{i=1}^{Z}{\bf R}_{pi}+g_{n}\Sigma_{i=1}^{A-Z}{\bf R}_{ni}|\psi>=
[1-g_{n}]<\phi|\Sigma_{i=1}^{Z}{\bf R}_{pi}|\psi>\eqno (5.1b)$$
  
\noindent (remembering $g_{p}\equiv 1$).  Thus, from Eqs. (3.2) and (5.1), 
the excitation rate $\Gamma$ in sec$^{-1}$ of e.g., 
one of the four valence protons from the ground state $|\psi>$ to one of
the 12 degenerate excited states $|\phi>$   
(any of the four protons may be excited, and the 1- state of this proton plus 
its subshell partner can have three possible orientations)  
can be expressed purely in terms of proton matrix elements:

$$\eqalign{\Gamma=&{\lambda\over 2a^{2}}[1-g_{n}]^2
12|<\phi|\Sigma_{i=1}^Z{\bf R}_{pi}|\psi>|^{2}\cr
=&{\lambda\over 2a^{2}}[1-g_{n}]^212\Sigma_{m}{4\pi\over 3} 
|<\phi|\Sigma_{i=1}^{i=Z} 
R_{pi}Y_{1,m}(\theta_{i}, \phi_{i})|\psi>|^{2}\cr}\eqno (5.2a,b)$$

\noindent A similar expression may be written for the excitation 
rate of a neutron totally in terms of neutron matrix elements.   

	It is worth remarking that the expression for the lifetime $\tau(\phi\rightarrow \psi)$ 
of one of the excited state $|\phi>$ to decay by an electric 
dipole transition to the ground state $|\psi>$ can be written in 
terms of the same matrix elements as appear in (5.2)$^{24}$:

$${1\over\tau(\phi\rightarrow \psi)}= {16\pi c\over 9}\Big({E\over \hbar c}\Big)^{3}
\Big({e^{2}\over \hbar c}\Big)\Sigma_{m} 
|<\phi|\Sigma_{i=1}^{i=Z} 
R_{pi}Y_{1,m}(\theta_{i}, \phi_{i})|\psi>|^{2}\eqno (5.3)$$

\noindent where $E$ is the energy difference of the two states.  
Thus we may express $\Gamma$ in terms of this lifetime:

$$\Gamma={\lambda\over 2a^{2}}[1-g_{n}]^2{9\over c(e^{2}/\hbar c)}
\Big({\hbar c\over E}\Big)^{3}{1\over\tau(\phi\rightarrow \psi)}\eqno (5.4)$$

\noindent Eq. (5.4) would be useful if we have the experimental lifetimes and 
branching ratios of the state $\phi$. Unfortunately, in this case we do not, 
so we are forced to estimate the matrix element in (5.2b). 

	In this paper we shall approximate the matrix element by using the  
same ``very rough estimate" employed by Blatt and Weisskopf$^{25}$ for calculating the 
lifetime $\tau$.  They assume the radial wavefunction of the proton in both states $\phi$ and 
$\psi$ is $\Theta (R_{0}-R)[3/R_{0}^{3}]^{1/2}$ where $\Theta$ is the step function and 
$R_{0}=1.4\times10^{-13}A^{1/3}$ is the nuclear radius.  As they point out, the actual 
radial integral is expected to be ``somewhat smaller" since radial wavefunctions oscillate: 
say, $\beta$ times smaller. We obtain from (5.2b) the result

$$\Gamma={\lambda\over a^{2}}[1-g_{n}]^2\beta^{2} R_{0}^{2}\eqno (5.5)$$

\noindent (a numerical factor 9/8 has been replace by 1). 	
	
	The expected total count ${\cal C}$ for an experimental run of $D$ kg-days from a transition 
due to a nuclide which comprises the fraction $X$ of the 
$8.3\times10^{24}$ atoms/kg in common Ge is found from 
(5.5), with the GRW parameters, to be 
 
 $${\cal C}={\beta^{2}\over 4}[1-g_{n}]^{2}XD\eqno (5.6)$$
 
\noindent For example, consider a transition in Ge$_{74}$ to the 2165 keV 1- state.   
We shall use the data from the Rico Grande experiment$^{22}$, similar to COSME but with 
$D=1.135$ kg-yrs=414.3 kg-days (COSME's $D=85.24$ kg-days) with $X=.14$  
(COSME's $X=.37$). Denoting by ${\cal C}_{\tenrm expt}$ the 
upper limit on the number of observed counts, we obtain
 
$$1+.26{{\cal C}^{1\over 2}_{\tenrm expt}\over \beta}\geq g_{n}
\geq 1-.26{{\cal C}^{1\over 2}_{\tenrm expt}\over \beta}\eqno (5.7)$$

\noindent For this experiment, the upper limit (obtained from the counts under a Chi$^{2}$ fit to 
a background quadratic plus the expected 
experimental resolution shape centered on 2165 keV) is 
${\cal C}_{\tenrm expt}=.89$ counts at the 68\% confidence level 
(3.9 counts at the 95\% confidence level). For
$\beta\approx .3$ one obtains $1.8\geq g_{n}\geq .2$.  However, our choice of $\beta$ 
is just hypothetical, as we have not made the effort to seriously evaluate the 
matrix element$^{26,27}$. The points to be made are that the range of $g_{n}$ is 
not so far from $g_{p}=1$ and that  
the various numbers involved in calculating (5.7) tantalizingly contrive to 
be on the edge of showing mass proportionality. 
Indeed, this would more easily be shown with a larger value of $\lambda/a^{2}$ than the GRW 
value: for instance, with $\lambda/a^{2}=100\lambda/a^{2}_{\tenrm GRW}$, 
the above inequality becomes $1.1\geq g_{n}\geq .9$. 
But, with the GRW parameters, it would require a  
long counting time and a proper calculation of 
matrix elements before one might say that $g_{n}/g_{p}\approx 1$.  
 
\vskip 1cm
\leftline {\title 6. Constraint on $g_{e}$/$g_{p}$.}
\bigskip

 	Here we apply Eq. (3.2) to calculate the spontaneous ionization rate of 
the 1s electrons in a Ge atom.  

	If a 1s electron is spontaneously ionized,  
the remaining electrons in the atom rapidly cascade into the (singly ionized) ground state, 
emitting a photon pulse of 11.1 keV (the ionization energy of a 1s electron).  The 
ionized electron also deposits its kinetic energy in the Ge sample, 
which augments the energy of the pulse. Thus the 
signature of these events is a distribution of photon pulses of energy $E>11.1$ keV.    

In reference 12, a Hartree calculation of the 
matrix element in (3.2) for the electrons was numerically performed, where $|\psi>$ is 
the ground state of Ge and $|\phi>$ is a state where a 1s electron is ionized.
The result of the calculation may be expressed as a function $C(E)$ which gives the expected 
pulse counting rate if GRW parameters are assumed and 
if the electron is totally responsible for collapse (i.e., $g_{e}=1$, $g_{n}=g_{p}=0$).  
$C(E)$ is zero for $E<11.1$ keV, abruptly rises to 5370 counts/keV/kg/day at $E=11.1$ keV, 
and decays in roughly exponential fashion, with the value $\approx 4000$ counts/keV/kg/day at 
$E=12$ keV, and $\approx 1500$ counts/keV/kg/day at $E=16$ keV.  
 
	If we assume that $g_{n}=g_{p}=1$ (as we shall hereafter do) then, as in the 
preceding section's Eq. (5.1), we can express the matrix element for the nucleons 
as $-M_{e}/M_{p}$ times the matrix 
element for the electrons. Putting this into Eq. (3.2), 
the resulting rate $\Gamma$ in counts/sec/kg/day may then be written as
	
	$$\Gamma={(\lambda/a^{2})\over (\lambda/a^{2})_{GRW}}
	[g_{e}-{M_{e}\over M_{p}}]^{2}C(E)\eqno (6.1)$$
	 
\noindent We note that the rate in (6.1) vanishes if 
there is mass proportionality ($g_{e}/g_{p}=M_{e}/M_{p}$). 

	 Figure 1 shows a graph of the counts/.1keV/85.24kg-day from COSME in 
the energy range 5 to 17 keV. A recent paper$^{21}$ describes a search with the same apparatus in   
a similar energy range.  These authors were looking at the 
TWIN data for a signature 11.1 keV 
peak resulting from a hypothesized violation of charge conservation (in which a K-shell electron 
decays to neutrals, and the other electrons in the atom readjust). They fit the data 
to three x-ray peaks (Cu, Zn and Ga) known to result from cosmogenic excitation of the Ge isotopes 
in the sample under observation, together with a background 
quadratic polynomial plus the hypothesized process, and they look at the 
difference between the fit and the data at the 68\% and 90\% 
confidence levels.  

	We employ here the same procedure.  Fig. 1 (solid line) shows the best 
fit to the data by this method, without the hypothesized process, 
a multiple of $C(E)$.  Superimposed upon this fit 
is a graph of $10^{-3}C(E)$ (dash-dotted line), folded in with the detector resolution shape  
(a gaussian of standard deviation .18 keV).  
It is clear that this is by no means a good fit to the data, so the coefficient of $C(E)$ 
in Eq. (6.1) is considerably smaller than $10^{-3}$.

	Also shown in Fig. 1 is the fit with the 
hypothesized process at $4.2\times 10^{-5}C(E)$ (dotted line) which corresponds to 
the 68\% confidence level. Not shown is a similar shaped curve at $9.7\times 10^{-5}C(E)$ 
which corresponds to the 95\% confidence level. 
 
	If we assume that the parameters $\lambda$ and $a$ are the same as those given 
by GRW, then we conclude from Eq.(6.1) that, at the 68\% confidence level, 
$[g_{e}/g_{p}-M_{e}/M_{p}]^{2}C(E)\leq 4.2\times 10^{-5}C(E)$ or

$$0\leq {g_{e}\over g_{p}}\leq 13{M_{e}\over M_{p}}\eqno (6.2)$$

\noindent  Thus, according to CSL with the GRW parameters, nucleons are 
mostly responsible for collapse.

	It is worth noting that it would take an improvement in the experimental limit by e.g., 
a factor of 1/300, which results in 
$.3M_{e}/M_{p}\leq g_{e}/g_{p}\leq 1.7 M_{e}/M_{p}$, to suggest that $g_{e}/g_{p}=0$ 
may be ruled out. However, it should also be noted that we need not be wedded to 
the GRW parameters.  Thus an increase of $\lambda/a^{2}$ by a factor of 300 
or more would have the same effect.  On the other hand it would take a decrease in $\lambda/a^{2}$ 
by a factor of $4\times 10^{-5}$ or more for the limit obtained in this experiment not to suggest that 
nucleons collapse more rapidly than electrons.

\bigskip
\centerline {\big Acknowledgments}
\bigskip

	We would like to thank Brian Collett for 
computational help.  One of us (P.P.) would like to thank the Institute for Advanced Studies 
of the Hebrew University in Jerusalem, 
where some of this work was done, for its support and hospitality, and
Yakir Aharonov for setting up the workshop at the Hebrew University.  
\bigskip

\vfill
\eject

\centerline {\big References}
\bigskip

\noindent 1. D. Bohm and J. Bub, Revs. Mod. Phys. {\bf 38}, 453 (1966).

\noindent 2.  P. Pearle, Phys. Rev. D{\bf 13}, 857 (1976). For more on this pre-GRW work, see
P. Pearle in {\it Sixty-Two Years of Uncertainty}, A. Miller ed. (Plenum, New York 1990), p. 193.

\noindent 3.  G. C. Ghirardi, A. Rimini and T. Weber, Physical Review D{\bf 34}, 470 (1986);
Physical Review D{\bf 36}, 3287 (1987); Foundations of Physics {\bf 18}, 1 (1988).

\noindent 4.  P. Pearle, Physical Review A{\bf 39}, 2277 (1989); G. C. Ghirardi, 
P. Pearle and A. Rimini, Physical Review A{\bf 42}, 78 (1990).    
For reviews see  G. C. Ghirardi and A. Rimini in {\it Sixty-Two Years of Uncertainty}, 
edited by A. Miller (Plenum, New  York 1990); 
G. C. Ghirardi and P. Pearle in {\it Proceedings of the Philosophy
of Science Foundation 1990, Volume 2}, edited 
by A. Fine, M. Forbes and L. Wessels (PSA Association, Michigan 1992), p. 19 and p. 35. 

\noindent 5. F. Karolyhazy, Nuovo Cimento {\bf42A}, 1506 (1966); 
P. Pearle and E. Squires, Found. Phys. {\bf 26}, 291 (1996). 

\noindent 6. P. Pearle, {\it Relativistic Collapse Model with Tachyonic Features} 
(Hamilton College preprint, 1997).

\noindent 7. J. R. Clauser (private communication).

\noindent 8. A. Zeilinger (private communication).

\noindent 9.  E. J. Squires, Phys. Lett. A{\bf 158}, 432 (1991).

\noindent 10. Q. Fu, Phys. Rev.A{\bf 56}, 1806 (1997).
 
\noindent 11.  P. Pearle and E. J. Squires, Phys. Rev. Lett. {\bf 73}, 1 (1994).

\noindent 12. B. Collett, P. Pearle, F. Avignone and S. Nussinov, Found. Phys. {\bf 25}, 1399 (1995).     
 
\noindent 13.  H. S. Miley, F. T. Avignone III, R. L. Brodzinski, J. I. Collar and 
J. H. Reeves, Phys. Rev. Lett. {\bf 65}, 3092 (1990).

\noindent 14.  
 R. Penrose in {\it Quantum Concepts in Space and Time}, 
edited by R. Penrose and C. J. Isham (Clarendon, Oxford 1986), p. 129;  
in {\it The Emperor's New Mind}, (Oxford University Press, Oxford, 1992); in {\it 
Shadows of the Mind}, (Oxford University Press, Oxford, 1994).
 
\noindent 15.  L. Diosi, Phys. Rev. A{\bf 40}, 1165 (1989); G. C. Ghirardi,
R. Grassi and A. Rimini, Phys. Rev. A{\bf 42}, 1057 (1990); ref. 6.

\noindent 16. P. Pearle, Foundations of Physics {\bf 12}, 249 (1982)

\noindent 17. We would like to thank Leo Stodolsky for bringing such experiments 
to our attention, and for providing information about the experimental accuracy.  
 
\noindent 18. J. I. Collar, PhD diss. U of SC 1992.
 
\noindent 19. E. Garcia et al Phys. Rev. D {\bf 51}, 1458 (1995).

\noindent 20. J. Morales et al, Nucl. Instr. Meth. A {\bf 321}, 410 (1992).

\noindent 21. A.K. Drukier et al, Nucl. Phys. B (Proc. Suppl.) {\bf 28A}, 293 (1992).

\noindent 22. Y. Aharonov et al Phys. Rev. D {\bf 52}, 3785 (1995).

\noindent 23. C. E. Aalseth et. al., Nuclear Phys. B (Proc. Suppl.) {\bf 48}, 223 (1996).

\noindent 24.  J. M. Blatt and V. F. Weisskopf, {\it Theoretical Nuclear Physics}, 
(Wiley, New York 1960), p. 595.

\noindent 25.  Ibid, p.625 et.seq.

\noindent 26.  R. D. Lawson, {\it Theory Of The Nuclear Shell Model}, 
(Clarendon, Oxford 1980).

\noindent 27. J. Joubert, F. J. W. Hahne, P. Navratil and H. B. Geyer,  
 Phys. Rev. C {\bf 50}, 177 (1994) and references therein.  

\vfill
\eject

\bigskip
\centerline {\big Figure Captions}
\bigskip

Figure. 1  A graph of the COSME data in the region 5-17 keV is shown along
with the best fit to the three known X-ray peaks plus a quadratic polynomial
background (solid curve).  Two additional curves are shown,  
corresponding to predicted rates folded in with the experimental resolution (a gaussian of width 
.18keV). 
The dash-dotted curve corresponds to $10^{-3}C(E)$  and the dotted curve to the 
68\% confidence level value of $4.2\times10^{-5}C(E)$.  

\bigskip
\vfill
\eject
\bye